\documentclass{PoS}
\usepackage[caption=false]{subfig}
\usepackage{commath}
\newcommand*\difff{\mathop{}\!\mathrm{d}}
\newcommand{\lambdabar}{{\mkern0.75mu\mathchar '26\mkern -9.75mu\lambda}}
\title{Cosmic ray ensembles from ultra-high energy photons propagating in the galactic and intergalactic space}

\ShortTitle{Cosmic ray ensembles from ultra-high energy photons}

\author{
Niraj Dhital$^{a}$,
Oleksandr Sushchov$^{a}$,
\speaker{Jan P{\k e}kala}$^{,a}$, 
Kevin Almeida Cheminant$^{a}$,
Dariusz G\'{o}ra$^{a}$,
Piotr Homola$^{a}$,
Alan R. Duffy$^{b}$,
Pawe\l{} Jagoda$^{a,c}$,
Robert Kaminski$^{a}$,
Marcin Kasztelan$^{d}$,
Peter Kovacs$^{e}$,  
Vahab Nazari$^{a,f}$,
Micha\l{} Nied\'zwiecki$^{g}$, 
Katarzyna Smelcerz$^{a,g}$,
Karel Smolek$^{h}$,
Krzysztof Rzecki$^{i}$, 
Jilberto Zamora-Saa$^{j}$,
Zolt\'{a}n Zimbor\'{a}s$^{e}$
\\
  
        E-mail: \email{niraj.dhital@ifj.edu.pl}\\
        $^{a}$ Institute of Nuclear Physics Polish Academy of Sciences, Radzikowskiego 152, Cracow, Poland\\
        $^{b}$ Centre for Astrophysics and Supercomputing, Swinburne University of Technology, Hawthorn, VIC 3122, Australia\\
        $^{c}$ AGH University of Science and Technology, 30-059 Cracow, Poland\\
        $^{d}$ National Centre for Nuclear Research, Andrzeja Soltana 7, 05-400 Otwock, Swierk, Poland\\    
        $^{e}$ Institute for Particle and Nuclear Physics,Wigner Research Centre for Physics, Hungarian Academy of Sciences, H-1525 Budapest, Hungary \\
        $^{f}$ Joint Institute for Nuclear Research, Dubna, Russia  \\
        $^{g}$ Institute of Telecomputing, Faculty of Physics, Mathematics and Computer Science, Cracow University of Technology, Warszawska 24st 31-155 Cracow, Poland\\
        $^{h}$ Institute of Experimental and Applied Physics, Czech Technical University in Prague\\
        $^{i}$ Cracow University of Technology, Warszawska 24, 31-155 Cracow, Poland \\
        $^{j}$ Universidad Andres Bello, Departamento de Ciencias Fisicas, Facultad de Ciencias Exactas, Avenida Republica 498, Santiago, Chile
        
        }

\abstract{Propagation of ultra-high energy photons in the galactic and intergalactic space gives rise to cascades comprising thousands of photons. Using Monte Carlo simulations, we investigate the development of such cascades in the solar magnetosphere, and find that the photons in the cascades are distributed over hundreds of kilometers as they arrive at the top of the Earth's atmosphere. We also perform similar study for cascades starting as far as 10 Mpc away from us using relevant magnetic field models. A few photons correlated in time are expected to arrive at the Earth from the latter type of cascade. We present our simulation results and discuss the prospects for detection of these cascades with the Cosmic-Ray Extremely Distributed Observatory.}

\FullConference{36th International Cosmic Ray Conference -ICRC2019-\\
		July 24th - August 1st, 2019\\
		Madison, WI, U.S.A.}

\begin{document}

\section{Introduction}
Although a definitive conclusion on the exact composition of ultra-high energy cosmic rays (UHECR) is lacking, the current observations and the conclusions drawn thereof strongly
suggest that UHECRs are hadronic in nature \cite{Abbasi:2018wlq, Aab:2014aea, Abbasi:2009nf}.
Analyses based on different observables measured from the extensive air showers (EASs) produced by the UHECRs suggest that the
primaries appear to be more like atomic nuclei heavier than protons \cite{Aab:2014aea}. In addition, it has also been established that some crucial features, e.g., the 
muon content in EASs produced by UHECRs, do not quite agree with what is obtained from the simulations of hadron-initiated showers.
More muons are present in EASs than the simulations suggest \cite{Aab:2016hkv}.
We also lack a definitive understanding of the mechanism that produces (accelerates) the primary particles at (to) such high energies. Of the two broad classes of models for
UHECR production or acceleration, the {\em bottom-up} models are based on assumption that the low energy particles are accelerated to the ultra-high energies (UHE)  
while the {\em top-down } 
models stem from assumption that the UHECRs are produced as a consequence of decay or annihilation of super-massive 
dark matter particles or topological defects\cite{Kachelriess:2008bk}.
If the former is solely responsible for UHECR production we should a see very small fraction of UHE photons, while if the latter is a dominant cause of UHECR production
we expect a much larger photon fraction. 

The most up-to-date results from the photon searches performed by
various experiments  show the non-observation of UHE photons and have placed 
stringent upper limits \cite{Abu-Zayyad:2013dii, Aab:2016agp}  disfavoring the top-down model of UHECR production. However, there are still compelling reasons for the UHE photon search.
The conclusions from the searches performed so far are based on the interaction models which are derived by extrapolating the experimental data measured at lower energies.
Eventually, the physics uncertainties grow significantly in the UHE regime, thus making it harder to correctly characterize the development of EASs. The mismatch between the muon number
from observations and that from simulation results for hadronic showers also appeals for a further investigation for UHE photon scenarios.
Both observation and non-observation of UHE photons will have a number of significant implications in fundamental science. It will help to establish or constrain better
several aspects of contemporary astrophysics e.g., models of dark matter, Lorentz invariance violation, space-time structures etc.

We present a method of UHE photon search that is based on electromagnetic cascading of UHE photons as they propagate through 
space. Such cascade development can occur when the primary photon traverses some region in space with magnetic field strong enough for pair-production, and subsequently
synchrotron emission from the produced electron-positron pair. As a consequence, a number of correlated particles or {\em cosmic ray ensemble}
(CRE)  from a UHE primary photon arrive at the Earth.
The number of particles and distribution of particles in CREs depend on the distance from the Earth to the region in space where cascading takes place, as well as the strength of
the magnetic field in the region.
\section{Simulations}
Using Monte Carlo simulations, we study UHE photon cascading for two different scenarios--- cascade development which occurs close to the Sun 
and that which occurs at much farther distance from the Earth. For the simulations of the first scenario, we use PRESHOWER 3.0 \cite{SPS_Sim}, while for the second scenario,
we use CRPropa 3 \cite{Batista:2016yrx}.
\subsection{PRESHOWER 3.0}
PRESHOWER 3.0 is the most recent version of simulation package PRESHOWER\cite{Homola:2003ru},
which was originally developed for the simulation of electromagnetic cascading of UHE photons in the geomagnetic
field. Treatment of most of the physics processes relevant for the cascading of UHE photon in PRESHOWER 3.0 has been adopted from \cite{Homola:2003ru}.
Some important features like a more accurate
particle tracking for 3D simulation and tracking of time is added in the new version of the code. This allows one to study spatial as well as temporal distribution of photons
which arrive at the top of the atmosphere. Also, PRESHOWER 3.0 allows one to study the {\em preshowering} of UHE photons not only in the geomagnetic field, but also in
the other regions in nearby space (e.g., close to the Sun).

A UHE photon subject to a magnetic field $H$ produces an $e^{+} e^{-}$ pair after traveling a path length $\difff l$ with a probability,
\begin{equation}
p_\mathrm{conv} = 1 - \exp\left(-\alpha \left(\chi\right) \difff l\right) \simeq \alpha \left(\chi\right)\difff l \mathrm{\ ,} \label{eqn:p_conv}
\end{equation}
where $\alpha $ is a function of variable $\chi \equiv \frac{1}{2} \frac{h \nu}{m_{\mathrm{e}} c^{2}} \frac{H}{H_\mathrm{cr}} $ with $H_\mathrm{cr} = 4.414 \times 10^{13} 
\ \mathrm{G}$ which can be approximated by,
\begin{equation}
\alpha\left(\chi\right) \approx \frac{0.16}{2 \chi} \frac{\alpha_\mathrm{em}}{\lambdabar_\mathrm{c}} \frac{H}{H_\mathrm{cr}} K_{1/3}^{2} \left(\frac{2}{3\chi}\right) \mathrm{\ ,}
\end{equation}
where $\alpha_\mathrm{em}$ is the fine structure constant, $\ \lambdabar_\mathrm{c}$ is the Compton wavelength of an electron and $K_{1/3}$ is a modified Bessel function.
Tracking of $e^{+} / e^{-}$ is performed using a modified Euler's method.

Also, the spectral distribution of energy of synchrotron photons emitted by the electrons is taken from \cite{Sokolov:1986nk},
\begin{eqnarray}
f(y) = \frac{9\sqrt{3}}{8 \pi} \frac{y}{( 1 + \xi y )^3}\left\{\int\limits_{y}^{\infty} K_{\frac{5}{3}}(z)\difff z 
+\frac{(\xi y)^{2}}{1 + \xi y}  K_\frac{2}{3} (y)\right\}\mathrm{\quad} \mathrm{\ ,}
\end{eqnarray}
where $y = \frac{ h \nu}{\xi \left( E - h \nu \right)} $ is the function of emitted photon energy $h\nu$, $\xi = \frac{3}{2} \frac{H_\perp}{H_\mathrm{cr}} \frac{E}{m_\mathrm{e} c^2}$, $E$ and $m_\mathrm{e}$ are energy and rest mass of electron respectively.
\begin{figure}%
    \centering
    \subfloat[Dipole field]{{\includegraphics[width=0.3\textwidth]{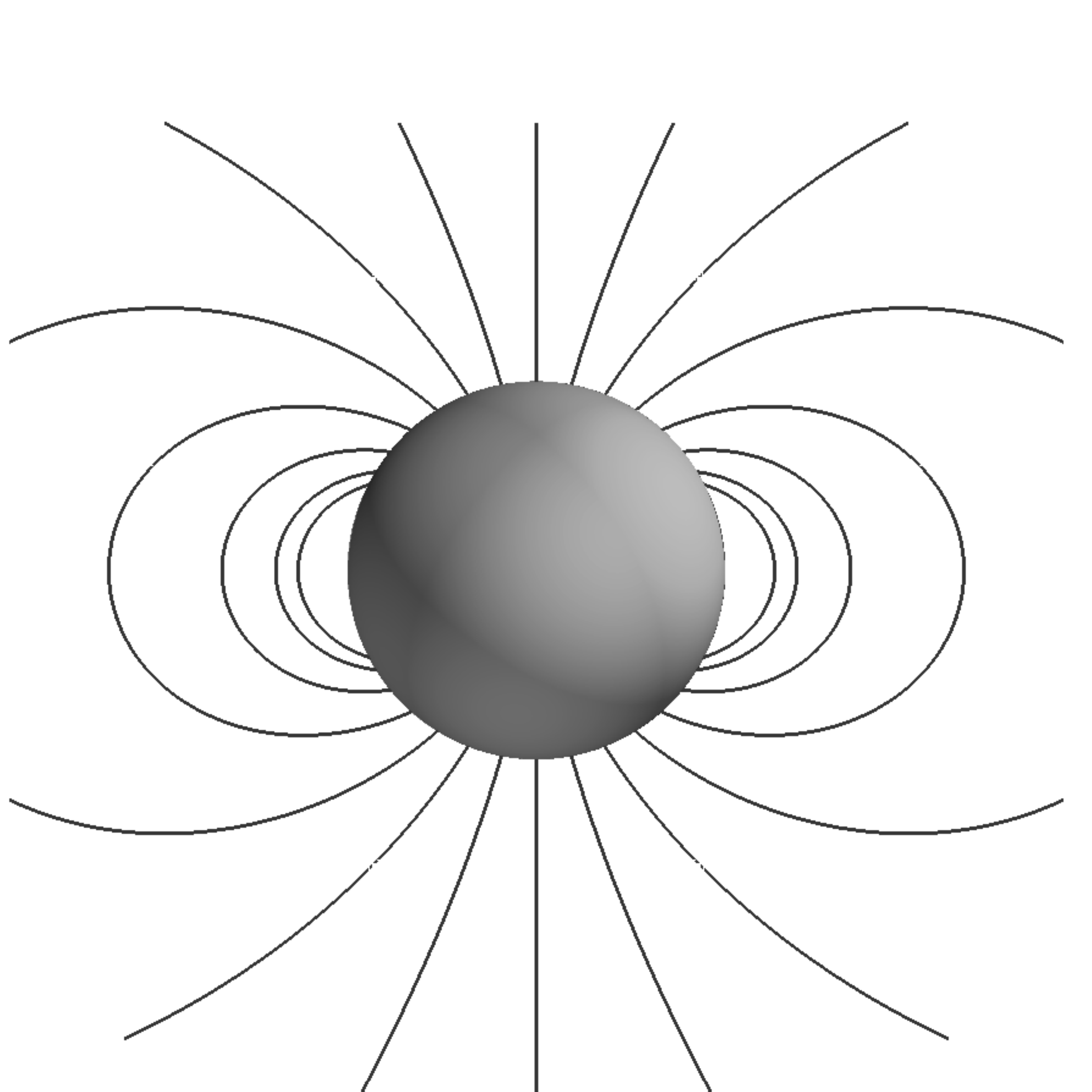}} }%
    \subfloat[DQCS field]{{\includegraphics[width=0.3\textwidth]{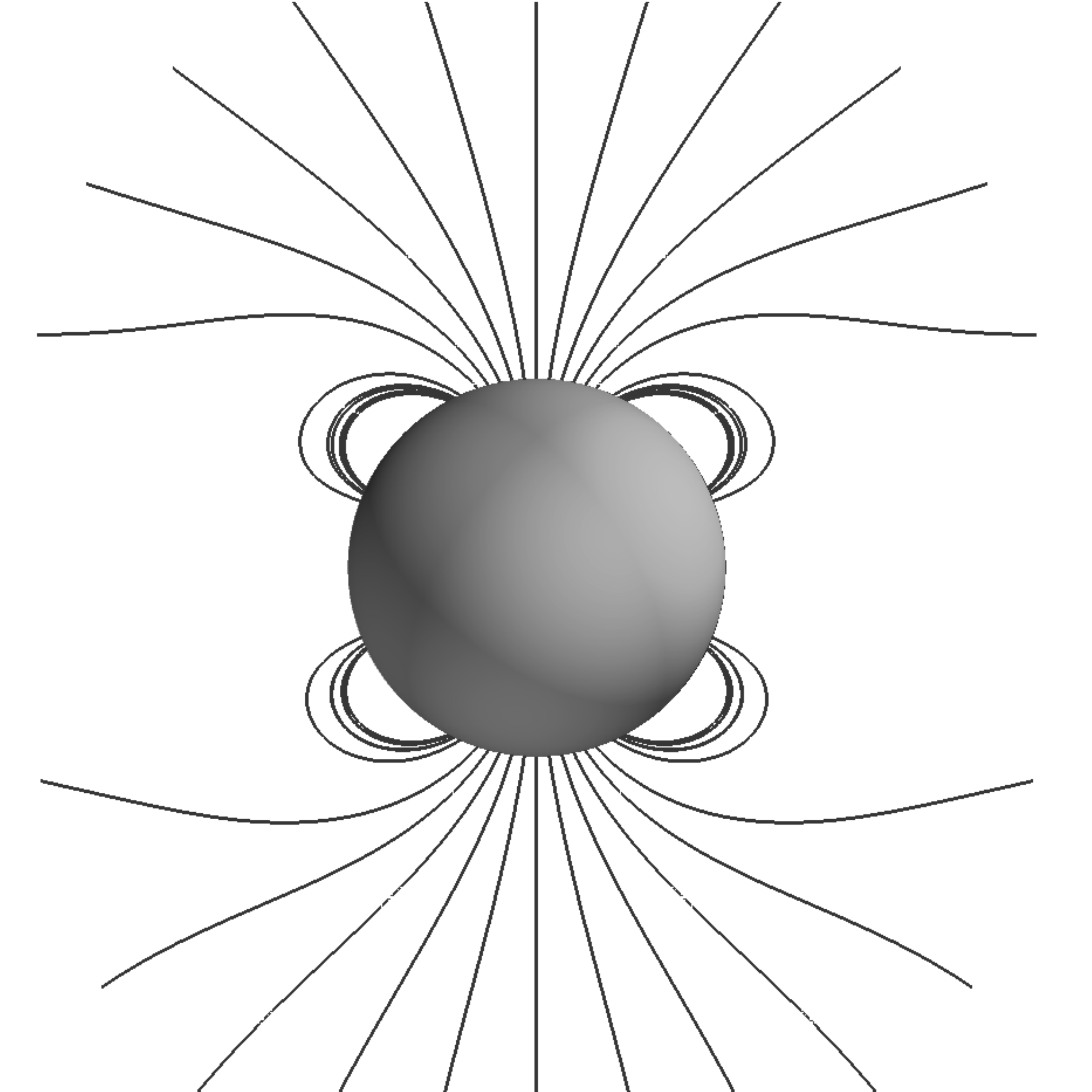} }}%
    \caption{Magnetic field models used in the simulation}%
    \label{fig:fields}%
\end{figure}

Two different models of solar magnetic field that we have used in the simulations of cascades developing close to the Sun are shown in Fig \ref{fig:fields}.
For the dipole model, the magnetic field is calculated considering a dipole with magnetic moment of $6.87 \times 10^{32}\  \mathrm{G\cdot cm^{3}}$. Although not a very
realistic model, using it for the solar magnetic field in our simulations allows us to study the effects of orientation of
considered dipole on the expected distribution of photons which arrive at the top of the Earth's atmosphere. The other model we used in our simulation 
is the so-called {\em dipole-quadrupole-current-sheet } (DQCS) model \cite{dqcs}, which gives a more realistic approximation of the solar magnetic field.
\subsection{CRPropa 3}
CRPropa 3 is a modular simulation framework designed for studying the propagation of ultra-relativistic particles through galactic as well as extragalactic space \cite{Batista:2016yrx}.
It allows for different spatial arrangements of sources while taking into account interactions of propagating particles with background
fields, their deflections in magnetic fields and cosmological effects. The modular structure provides
users with a wide range of simulation options, either by combining existing modules or adding their own
Python or C++ modules.

The propagation of the injected particles (each having its starting parameters specified) is
performed in discrete steps until the breaking conditions (energy threshold or reaching
the observer) are fulfilled. The probability for an interaction is calculated within each propagation step
with the use of the interaction length $\lambda$ of a specific process.
To make the calculations computationally inexpensive, interaction rates are calculated separately for
each process beforehand and implemented in the code during runtime.
\section{Results}
We summarize our results from the simulation of electromagnetic cascading of UHE photons propagating in space with regions characterized by some magnetic field.
In the case of a cascade initiated close to the Sun, there are several thousand low energy synchrotron photons which are practically aligned along a line
as they arrive at the top of the Earth's atmosphere. Although synchrotron emission from the electrons in the solar magnetic field (with a typical strength of around a few Gauss)
is likely even at much lower energies, the pair-production of $e^{+} e^{-}$ for a primary UHE photon is highly constrained by its primary energy.
\begin{figure}
\centering
    \includegraphics[width = 0.7\textwidth]{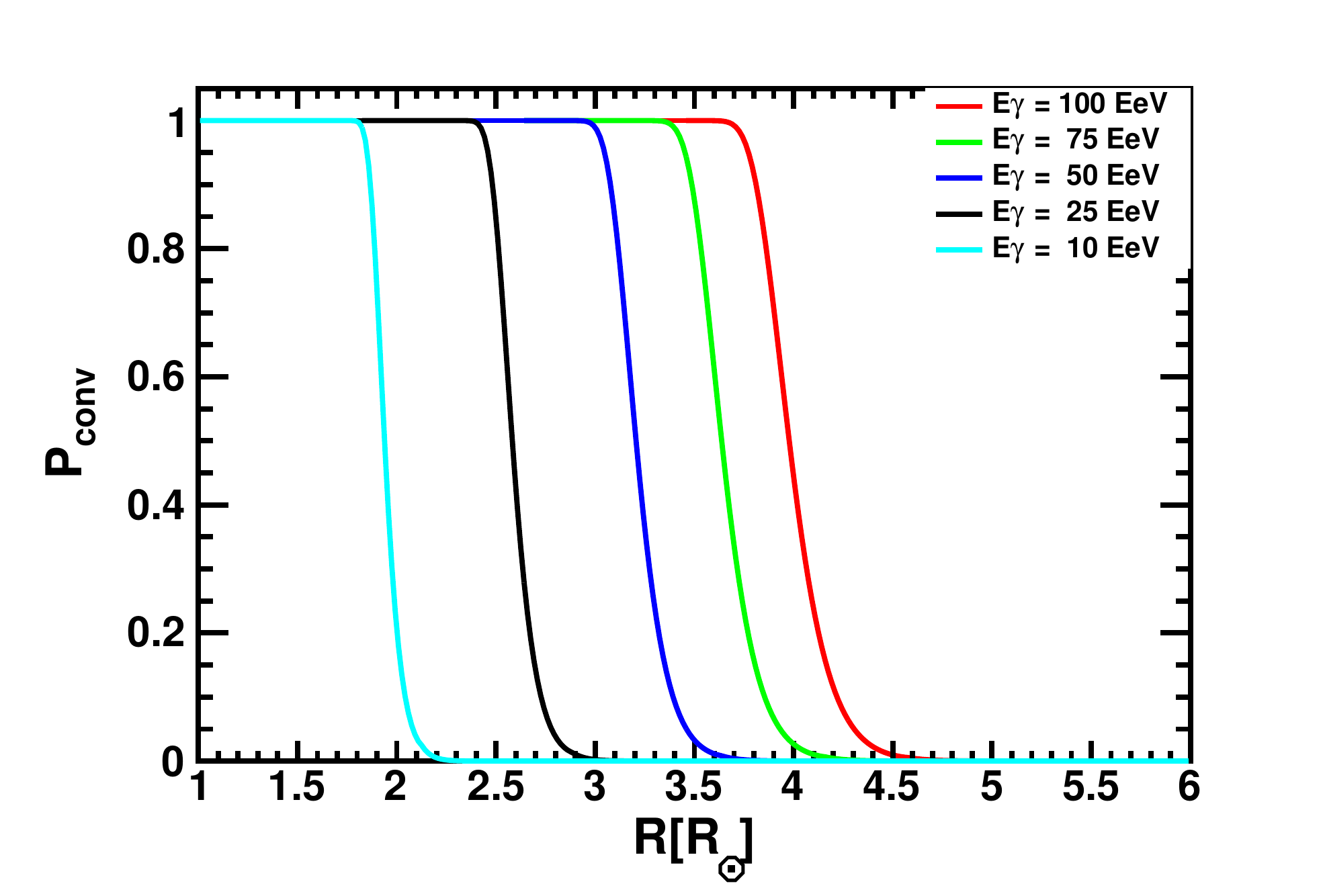}
\caption{Probability of magnetic pair production ($\gamma \rightarrow e^{+} e^{-}$) as a function of the impact parameter for UHE photons heading towards the Earth from the Sun's vicinity.
} \label{fig:conversion_p}
\end{figure}
In Fig. \ref{fig:conversion_p}, the magnetic pair production probability as a function of impact parameter of primary photon with respect to the Sun is shown for different energies. 
\begin{figure}
\centering
    \includegraphics[width = 0.5\textwidth]{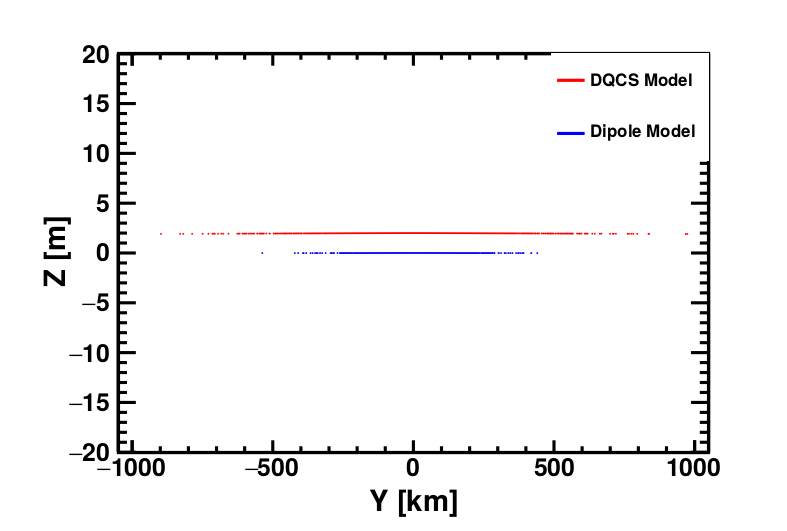}%
    \includegraphics[width = 0.5\textwidth]{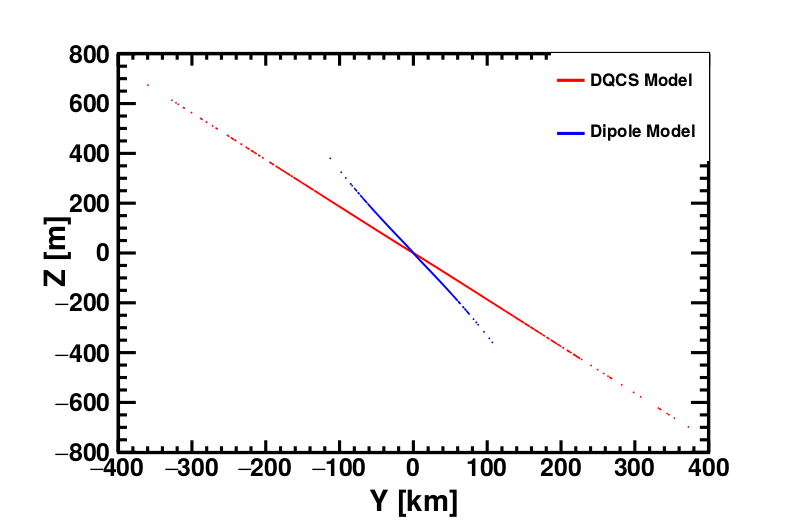}
\caption{Spatial distribution of photons with energies $> 10^{12} \mathrm{\ eV}$ arriving at the top of the atmosphere for a CRE produced by 100 EeV photon. 
The primary photon is directed towards the Earth such that the position of the closest approach has heliocentric latitude $90^\circ$(left panel) and $45^\circ$(right panel).
In the left panel, the distribution shown in red is shifted by 2 km in the positive $z$ direction.
}\label{fig:signature}
\end{figure}
In Fig. \ref{fig:signature}, typical signatures of CREs produced by a UHE photon heading towards the Earth via close vicinity of the Sun are shown. The spatial extent of the photons
in these CREs is several hundred kilometers. The signatures look different for different models of solar magnetic field, however, the {\em line}-signature is preserved in either of the
models used.
\begin{figure}%
    \centering
    \includegraphics[width = 0.5\textwidth]{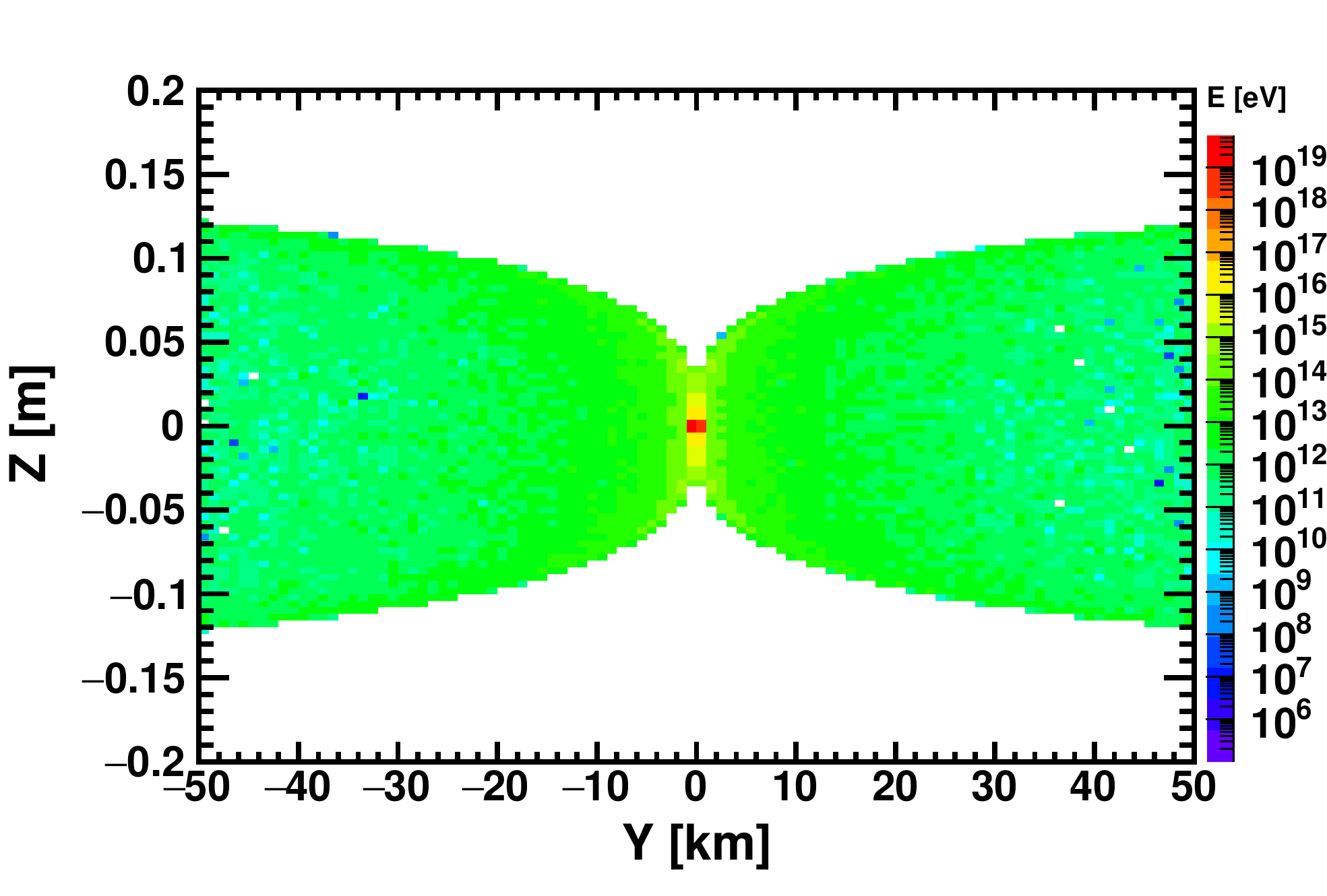}%
    \includegraphics[width = 0.5\textwidth]{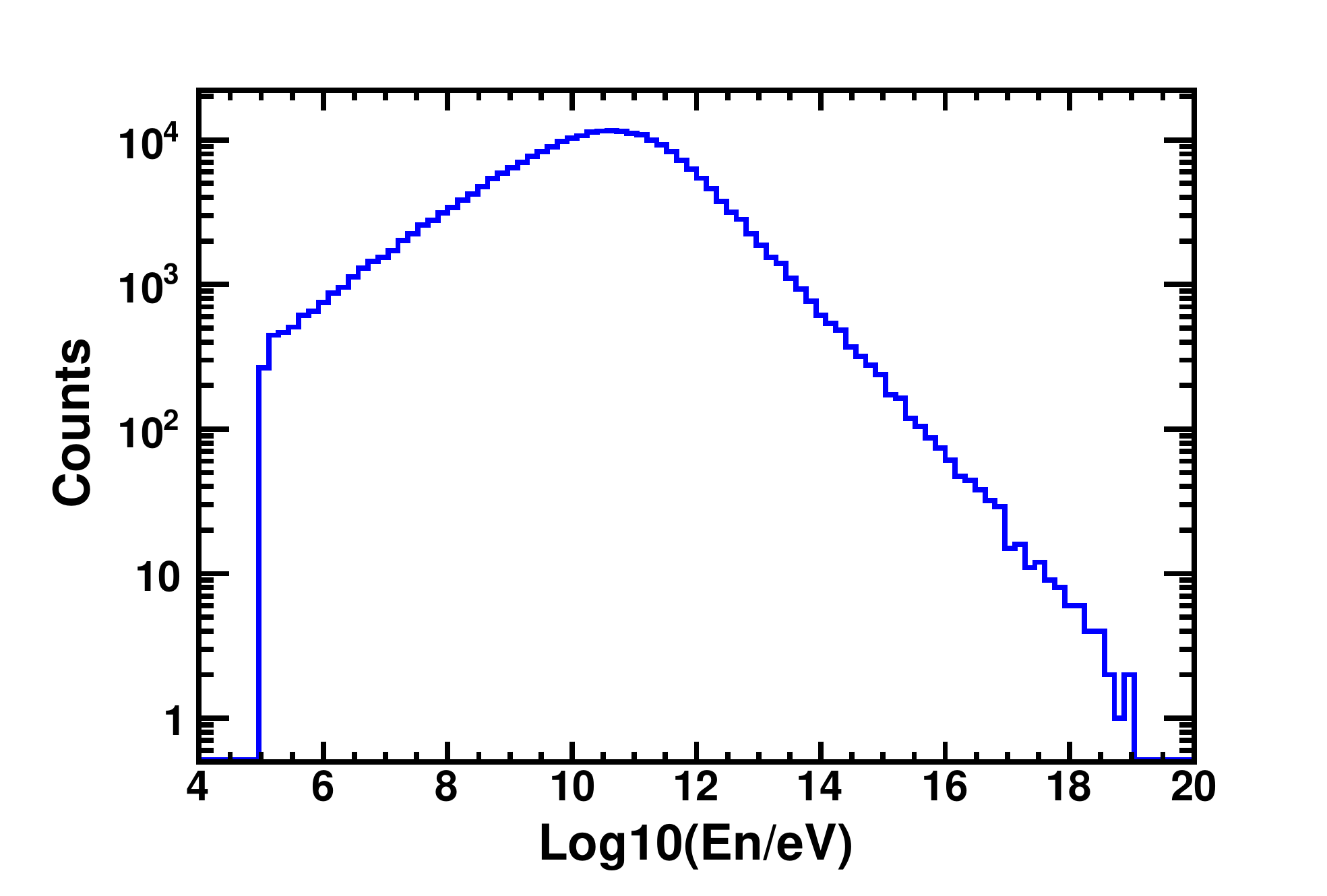}
    \caption{Distribution of energy of CRE photons arriving at the top of the atmosphere for a CRE produced by 100 EeV photon. The primary photon is directed towards the Earth such 
    that the position of the closest approach has heliocentric latitude $0^\circ$, and its impact parameter is $3 R_\mathrm{\odot}$.
    Note the difference in the scales along $x$ and $y$ axes. Energy distribution of CRE photons with energies larger than  $10^{5} \mathrm{eV}$ for the same CRE.} \label{fig:synch}
\end{figure}
Left panel in Fig. \ref{fig:synch} displays the spatial energy distribution of photons for the CRE in the left panel of Fig. \ref{fig:signature}. The most energetic photons are present 
near the {\em core} of the CRE. On the right panel of the figure, energy distribution of the photons in the same CRE is shown.

\begin{figure}%
    \centering
    \includegraphics[width = 0.45\textwidth]{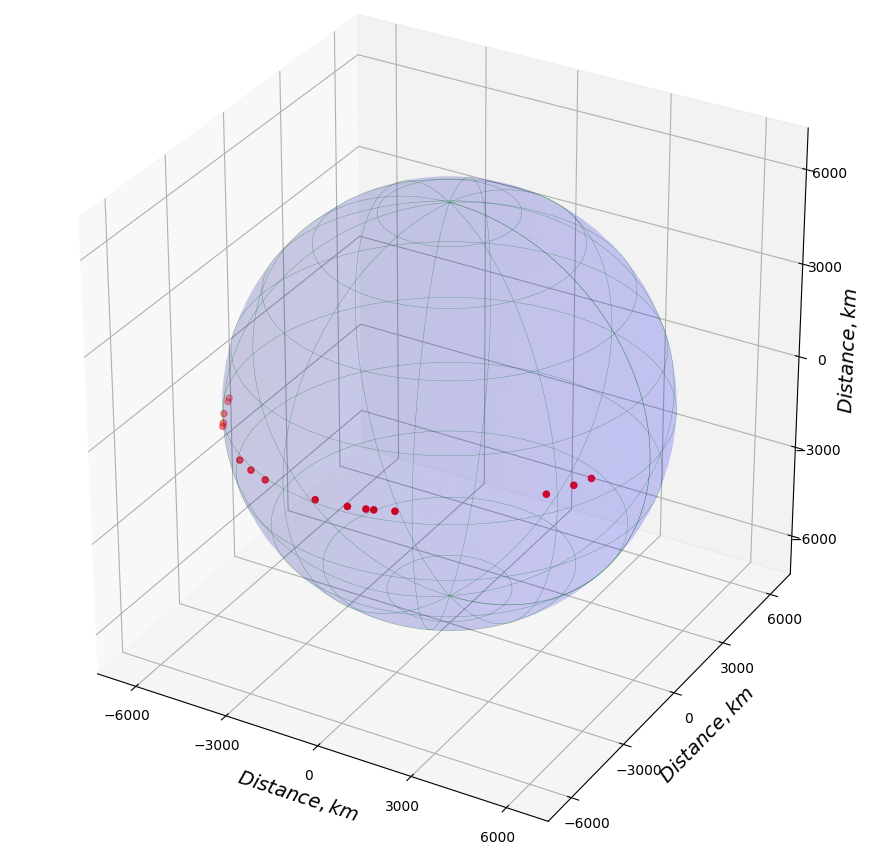}%
    \includegraphics[width = 0.55\textwidth]{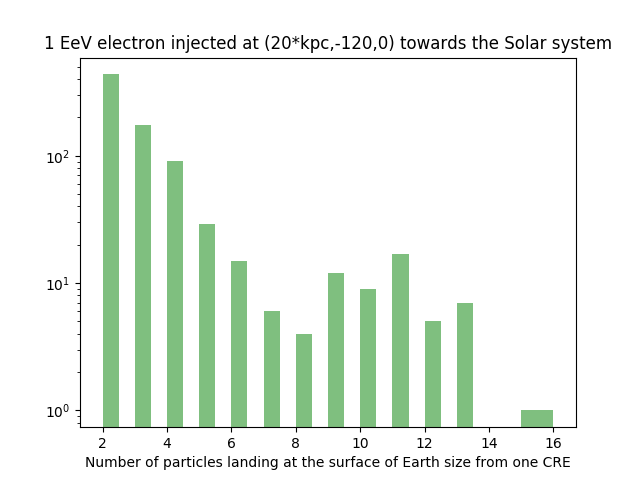}
    \caption{Left panel: Positions of CRE photons arriving at the Earth. Right panel: Distribution of number of photons with energies $> 1$ TeV landing at the Earth. CRE in both cases is 
    produced by a 1 EeV electron entering the galaxy and heading towards the Solar system.} \label{fig:crpropa}
\end{figure}
In case of a CRE originating at a large distance from the Earth we skip the first $e^{+} e^{-}$ pair
production and simulate EeV electrons entering the galaxy, considering Jansson \& Farrar galactic magnetic field model \cite{Jansson:2012pc}.
We study the distribution of resulting photons on the observational sphere with its center at the
electron's injection point, containing the Solar system on its surface. 
We then calculate the number of
photons, reaching the surface of an Earth-sized sphere in the Solar system, thus simulating the arrival of CREs at the Earth. 
In the left panel of Fig. \ref{fig:crpropa}, an example distribution of particles on the test sphere of the Earth size is
shown. In this particular example, as many as 16 photons make their way to the Earth.
In the right panel, distribution of number of photons belonging to the same CRE, capable to reach the Earth ``together'' is shown.
\section{Summary}
If UHE photons propagate towards the Earth from their sources, they are subjected to several processes. We demonstrated through simulations that they cascade on their way 
while traversing regions in space with sufficient magnetic field. Instead of a
single primary UHE photon, there is a possibility that a cascade of several correlated particles from a photon primary arrives at the Earth.
Results from simulations performed for UHE photons traveling through the Sun's vicinity show that a large number of photons ($\sim 10^{4}$) are produced.
Photons in these CREs are spatially extended over a distance of several hundred kilometers as the CRE arrives at the top of the Earth's atmosphere and are distributed
practically along a line. Also, they span a wide range of energy --- from GeV to EeV.

CREs which originate at large distances can span the whole Earth, and as in the other scenario we studied, its signature is also a very extended line.
However, a straight-forward detection of such signature is limited by relatively lower number of particles reaching the Earth.

CREs provide us with a unique opportunity to explore UHE photons. Contrary to the currently used common approaches which are based on observables measured 
in EASs produced by primary UHECRs, an alternative approach for photon search based on CREs can be used.
The new approach will enable us to detect UHE photons which {\em preshower} before arriving at the Earth's atmosphere, which is difficult to do by implementing
currently used common methods.

The main objective of Cosmic-Ray Extremely Distributed Observatory (CREDO) \cite{credo} is to study cosmic rays with a particular emphasis on CREs.
CREs produced by UHE photons at regions close to the Sun give rise to a line-like footprint several hundred
kilometers long at the top of the atmosphere, and consequently similar footprint of secondary particles is expected at the ground. For CREs produced at very far distances,
we expect a few tens of correlated particles that arrive at the Earth. CREDO, which is a global network of cosmic ray detectors, will enable us to detect CREs produced in either of the
scenarios.
\acknowledgments
This research has been supported in part by PLGrid Infrastructure. We warmly thank the staff at ACC
Cyfronet AGH-UST, for their always helpful supercomputing support. CREDO mobile application was developed in Cracow University of Technology.

\end{document}